\documentclass{amsart}

\usepackage{makecell}
\usepackage{url}
\usepackage[utf8]{inputenc}
\usepackage{indentfirst}
\usepackage{natbib}
\usepackage{graphicx}
\usepackage{float}

\newcommand{\notarxiv}[1]{}
\newcommand{\arxiv}[1]{#1}

\newcommand{\iqtree}{IQ-TREE}
\newcommand{\treetime}{TreeTime}

\newcommand{\beginsupplement}{%
        \setcounter{table}{0}
        \renewcommand{\thetable}{S\arabic{table}}%
        \setcounter{figure}{0}
        \renewcommand{\thefigure}{S\arabic{figure}}%
}

\title[Topology Fixation in Phylodynamic Inference]{Assessing the Validity of the Fixed Tree Topology Assumption in Phylodynamic Inference}

\begin{document}

\author[Fourment et al.]{
Mathieu Fourment,$^{\ast,1}$
Jiansi Gao,$^{2}$
Marc A Suchard$^{3,4,5}$
Frederick A Matsen IV$^{6,7,8,9}$
}

\maketitle

\noindent{\small\it
$^{1}$~Australian Institute for Microbiology and Infection, University of Technology Sydney, Ultimo, 2007, Australia\\
$^{2}$~Computational Biology Program, Fred Hutchinson Cancer Center, Seattle, 98109, USA\\
$^{3}$~Department of Human Genetics, University of California, Los Angeles, 90095, USA\\
$^{4}$~Department of Computational Medicine, University of California, Los Angeles, 90095, USA\\
$^{5}$~Department of Biostatistics, University of California, Los Angeles, 90095, USA\\
$^{6}$~Public Health Sciences Division, Fred Hutchinson Cancer Research Center, Seattle, 98109, USA\\
$^{7}$~Department of Statistics, University of Washington, Seattle, 98109, USA\\
$^{8}$~Department of Genome Sciences, University of Washington, Seattle, 98109, USA\\
$^{9}$~Howard Hughes Medical Institute, Fred Hutchinson Cancer Research Center, Seattle, 98109, USA\\
}

\medskip
\noindent{\bf Corresponding author:}
Mathieu Fourment, Australian Institute for Microbiology and Infection, University of Technology Sydney, Ultimo NSW, Australia, E-mail: mathieu.fourment@uts.edu.au

\begin{abstract}
Fixed tree topologies are widely used in phylodynamic analyses to reduce computational burden, yet the consequences of this assumption remain insufficiently understood.
Here, we systematically assess the impact of various fixed-topology strategies on phylogenetic and phylodynamic parameter estimates across a diverse set of viral datasets.
We compare fully Bayesian joint inference with fixed-topology strategies, including conditioning on maximum likelihood trees subsequently dated with LSD or \treetime.
Our analyses show that global parameters of the substitution and site models are largely robust to the fixed-topology assumption, whereas parameters that depend on the temporal structure of the tree, such as molecular clock rates, node ages, and demographic histories, can exhibit substantial biases.
We do treat unconstrained Bayesian analyses as the reference, although we recognize that these too are model-based approximations.
Nevertheless, our results highlight serious discordance associated with fixing the topology and underscore the need for faster, time-aware methods that simultaneously integrate topology and parameter estimation.
These findings raise important questions about the balance between computational efficiency and inferential accuracy in phylodynamic studies.
\end{abstract}

\arxiv{
}
\notarxiv{
\bigskip
}

\arxiv{
\section{Introduction}
Phylodynamics is an interdisciplinary field that combines evolutionary biology, epidemiology, and genomics to understand the spread and dynamics of infectious diseases through time.
By analyzing genetic sequences of pathogens, phylodynamic methods allow researchers to reconstruct transmission histories \citep{stadler2011mammalian}, estimate key epidemiological parameters (such as reproduction numbers \citep{stadler2013birth} and population size changes \citep{minin2008smooth}), and identify the origins and spread routes of outbreaks \citep{lemey2010phylogeography}.
This approach has proven especially valuable during rapidly evolving epidemics, providing near real-time insights that inform public health responses \citep{dellicour2018phylodynamic,hadfield2018nexstrain,lemey2021untangling}.

The phylogenetic tree is a central object in phylodynamic analysis, encoding the evolutionary relationships among sampled sequences.
This tree has two key components: the topology, which describes the branching structure (i.e., the pattern of how lineages split and relate to each other), and the branch lengths, which represent the amount of genetic change or elapsed time along each branch.
Accurate inference of both components is critical for meaningful phylodynamic interpretation.

Bayesian inference methods, such as those implemented in the sister software packages BEAST X \citep{baele2025beast} and BEAST 2 \citep{bouckaert2019beast}, provide a powerful framework for phylodynamic analysis by allowing the joint estimation of evolutionary, demographic, and epidemiological parameters from genetic sequence data.
Bayesian methods integrate over uncertainty in model parameters, including tree topologies, node heights/branch lengths, molecular clock rates, and population size trajectories, yielding a coherent and comprehensive picture of pathogen evolution.
Bayesian methods have been widely adopted for their ability to model these intricate biological processes in a statistically coherent way, enabling researchers to extract rich temporal and spatial insights from genomic data and metadata \citep{hassler2023data}, especially during emerging epidemics.

Despite the flexibility and rigor of full Bayesian inference, its computational cost can be prohibitive, particularly for large datasets or time-sensitive analyses during outbreaks.
To improve computational speed, a range of approximate methods have been developed that retain some connection to Bayesian inference while assuming that the tree topology is known and fixed \citep{guindon2010bayesian,reis2011approximate,demotte2025iq2mc}.
Heuristic approximations, often involving multi-step procedures but offering substantial gains in speed, include maximum likelihood methods such as \treetime\ \citep{sagulenko2018treetime}, least-squares dating tools like LSD \citep{to2016lsd}, and other distance-based approaches \citep{tamura2012estimating,volz2017scalable}.
The fixed-tree assumption is particularly prevalent in variational inference frameworks, where it enables scalable approximations to the posterior \citep{fourment2014physher,fourment2019phylostan,swanepoel2022treeflow,fourment2025torchtree}.
A common feature among these tools is the assumption that the phylogenetic tree topology is known and fixed, which greatly simplifies calculations and enables the use of efficient algorithms.
While these methods are often sufficient for basic molecular clock dating and exploratory analyses, their reliance on a fixed topology can introduce biases when topological uncertainty is significant, especially for downstream inference of dynamic processes like population size changes or transmission patterns.

In this study, we systematically evaluate whether fixing the tree topology impacts the accuracy of phylodynamic parameter estimates using the Bayesian framework.
Specifically, we ask how biased the resulting estimates are when the topology is treated as known: is the posterior mean still accurate while the variance is simply underestimated for important, continuous phylogenetic or phylodynamic parameters of interest, as intuition might suggest?
Or does fixing the topology lead to fundamentally incorrect inference across the board?
Do the answers to these questions depend on how the fixed topology is obtained?
To answer these questions, we reanalyze a collection of previously published viral genomic datasets that were originally studied using BEAST X, which samples from the full posterior distribution simultaneously over trees \citep{Gao2025-bf}.
We treat these full Bayesian results as the reference, or `ground truth', against which we measure differences arising from the fixed-topology assumption.
By comparing the full Bayesian results to those obtained under various fixed-topology scenarios, we aim to quantify the consequences of this common approximation and assess the validity of fixed-topology methods for different types of phylodynamic inference.
}

\section{Materials and Methods}

We reanalyzed a set of empirical datasets that were previously studied \citep{Gao2025-bf} using BEAST X, which samples from the full posterior distribution over tree space and all other model parameters.
We refer to these analyses as \textit{unconstrained} analyses.
For each dataset, we conducted additional BEAST analyses in which the rooted tree topology was fixed while keeping the rest of the model identically-structured and random; we refer to these as \textit{fixed-topology} analyses.
Fixed rooted topologies were obtained using two different approaches.

The first approach follows a commonly used two-step procedure: a maximum likelihood tree is first inferred without a molecular clock, and a separate, heuristic method is then used to estimate the root position, clock rate, and node ages.
In this study, we used \iqtree\ to infer the rate-free maximum likelihood (ML) tree, followed by divergence time estimation using either LSD or \treetime.
It is important to note that we use LSD or \treetime\ to select the root position, and the divergence times and rate estimates are discarded.
The focus of this study is the effect of using a fixed tree on parameter estimates, not benchmarking LSD and \treetime.
\treetime\ is integrated into the Nextstrain platform \citep{hadfield2018nexstrain}, which is widely used for real-time outbreak monitoring, while LSD is a widely adopted method for divergence time estimation in research studies \citep{dhanasekaran2022human}, particularly suited for large phylogenies.

The second approach selects a fixed topology from trees sampled in an unconstrained BEAST analysis.
While not practical for routine use, this gives an upper bound on the accuracy of a fixed-topology analysis.
Specifically, we used either the maximum clade credibility (MCC) tree or the sampled tree with the highest posterior probability.
Additionally, we explored an intermediate strategy: we rooted the ML topology inferred by \iqtree\ using what we call the maximum rooting credibility (MRC) method.
Specifically, we computed the frequency of each root-induced split across the BEAST posterior sample and selected the most frequent one that was also compatible with the ML topology.
This rooting corresponds to the most credible placement of the root among those supported by both the posterior and the ML tree.
Only five of the 15 datasets we analyzed met this strict compatibility criterion; for the remaining ten datasets, the corresponding ML topology did not contain any of the root splits sampled from their respective unconstrained posterior distributions.
This approach bypasses LSD or \treetime\ and allows us to assess whether the rate-free ML topology is compatible with a time-tree analysis.
It is important to note that in the fixed-topology analyses, only the tree topology (including the root position) is fixed, while the node ages are still estimated from the data.

Each analysis estimated the same set of parameters, including node ages, substitution model parameters (e.g., GTR rates and base frequencies), clock rates, and demographic parameters in coalescent models.
In this paper we classify parameters according to the BEAST model structure.
Parameters associated with the \textit{substitution model} include the rate bias and nucleotide frequency parameters of the GTR or HKY substitution models.
Parameters related to the \textit{site model} include the proportion of invariant sites, the shape parameter of the gamma distribution, and the relative rate parameters in unlinked models.
The \textit{tree model} pertains to node heights, most notably the root height.
The \textit{coalescent model} encompasses the effective population size parameters and the precision parameter of the Gaussian Markov random field prior \citep{minin2008smooth} in a piecewise-constant population size model.

We compared these estimates to those obtained from the unconstrained analyses, treating the latter as ground truth.
While no empirical analysis provides absolute truth, the unconstrained approach accounts for uncertainty in both the topology and model parameters, making it the most comprehensive and thus the best available reference point.

For the piecewise-constant population size parameters, we calculated the root mean squared relative error (RMSRE) of their posterior mean estimates
$$
\operatorname{RMSRE}(\mathbf{\bar{x}}) = \sqrt{\frac{1}{n} \sum_{i=1}^n \left( \frac{\bar{x}_i - \bar{x}_i^u}{\bar{x}_i^u}\right)^2}
$$
where $\bar{x}_i$ and $\bar{x}_i^u$ are the mean population size of epoch $i$ estimated with the fixed-topology and unconstrained analyses, respectively.
For other parameters, such as the root height and substitution rate, we calculated the relative bias:
\[
\delta_x = \frac{\bar{x} - \bar{x}^u}{\bar{x}^u}.
\]

To quantify the topological differences between phylogenetic trees inferred by different models and methods, we computed pairwise Robinson-Foulds (RF) distances among a set of tree topologies sampled from BEAST (time tree posterior sample) and the maximum likelihood unrooted tree topology inferred by \iqtree.
All BEAST-sampled trees were de-rooted prior to comparison to ensure that the RF metric, which is defined for unrooted trees, was applicable.
The RF metric was used to compare the degree of topological disagreement between each pair of trees, without considering branch lengths.
The pairwise RF distances were normalized by dividing each distance by the maximum observed distance across all pairs, and subtracting the result from 1, such that trees with greater topological similarity have larger proximity values:

\[
d'_{ij} = 1 - \frac{\mathrm{RF}(T_i, T_j)}{\max\limits_{k,l} \mathrm{RF}(T_k, T_l)} .
\]
This transformed similarity matrix was then used to generate a spatial layout of the trees using a force-directed graph drawing algorithm \citep{fruchterman1991graph}, in which attractive and repulsive forces between nodes reflect their relative topological similarity.

For each dataset, we defined the mean ML--posterior distance as

\[
\bar{d}_{\text{ML}} = \frac{1}{N} \sum_{j=1}^{N} \mathrm{RF}(T_{\text{ML}}, T_j),
\]
where $T_{\text{ML}}$ is the maximum likelihood topology, $T_j$ is the $j$-th tree sampled from the unconstrained BEAST posterior, and $N$ is the number of posterior trees.
To compare values across datasets, we standardized $\bar{d}_{\text{ML}}$ using z-scores:

\[
z_i = \frac{\bar{d}_{\text{ML}, i} - \mu_i}{\sigma_i},
\]
with $\mu_i$ and $\sigma_i$ denoting the mean and standard deviation of distances between BEAST posterior trees for dataset $i$.
In this framework, $z \approx 0$ indicates that the ML topology is about as close to the BEAST posterior as the average dataset,
$z < 0$ indicates greater similarity than average, and $z > 0$ indicates greater dissimilarity.

All datasets, analyses, and comparisons were performed using a reproducible open-source pipeline implemented in Nextflow, available at: \\ \centerline{\url{https://github.com/4ment/fixed-tree-experiments}.}

\subsection{Datasets}
The datasets used in this study span a variety of viral pathogens and sampling contexts (Table~\ref{tab:data}).
The alignments range in size from 297 to 4,007 sequences, with sequence lengths spanning from 438 to 29,409 base pairs.
These datasets include: Ebola virus (EBOV) \citep{dudas2017virus,mbala2021ebola}, Influenza A and B viruses (IAV, IBV) \citep{worobey2014synchronized,bedford2015global}, HIV \citep{faria2014early}, Lassa virus (LASF) \citep{klitting2022predicting}, Mumps virus (MuV) \citep{moncla2021repeated}, Rabies virus (RABV) \citep{viana2023effects}, SARS-CoV-2 \citep{candido2020evolution,lemey2021untangling,pekar2022molecular}, West Nile virus (WNV) \citep{dellicour2020epidemiological}, and Zika virus (ZIKV) \citep{grubaugh2017genomic}.

The models used in this study follow a common structure composed of three key components: a coalescent prior, a substitution model, and a molecular clock model.
For the coalescent prior, we used either a constant population size model or a nonparametric Skygrid model \citep{gill2013improving} that allows for flexible demographic changes over time.
Nucleotide substitution models were either Hasegawa-Kishino-Yano (HKY) or general time reversible (GTR), with some analyses incorporating a discretized gamma distribution (4 categories) or a proportion of invariant sites to account for rate heterogeneity across sites.
To model evolutionary rate along lineages, we applied one of three molecular clock models depending on the dataset: a strict clock assuming a single rate across the tree, a relaxed uncorrelated log-normal (UCLN) clock \citep{drummond2006relaxed}, or a set of local clocks allowing rate shifts between lineages \citep{yoder2000estimation}.

\section{Results}

As described above, we reanalyzed several empirical datasets with BEAST X under two scenarios: unconstrained, sampling both topology and continuous parameters, and fixed-topology, where only the continuous parameters were estimated while the rooted tree was held constant.
Fixed topologies were obtained either from a rate-free ML tree rooted with \treetime\ or LSD, or from a summary tree derived from posterior samples of an unconstrained BEAST analysis (including the maximum clade credibility and maximum sampled posterior trees).

\begin{figure}[H]
\centering
\includegraphics[width=0.95\textwidth]{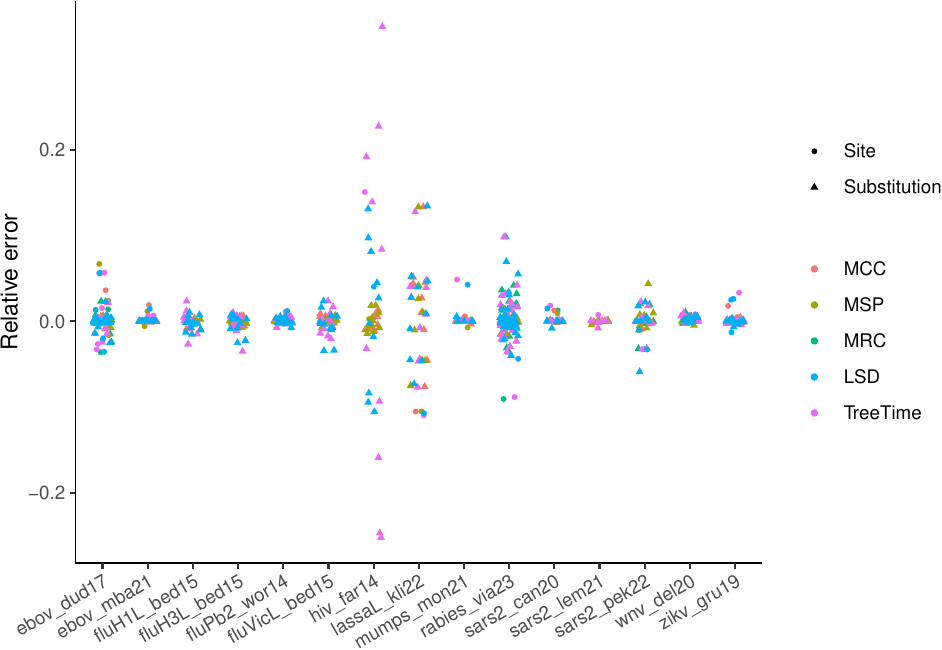}
\caption{\
Site and substitution model parameters are relatively insensitive to various fixed-tree assumptions.
Results are shown for five fixed-topology methods: maximum clade credibility (MCC), maximum sampled posterior (MSP), maximum rooting credibility (MRC), LSD, and \treetime.
}%
\label{fig:insensitive}
\end{figure}

\textbf{Substitution and site model parameter estimates are robust to the fixed-topology assumption.}
We first asked which model parameters are relatively insensitive to the fixed-topology constraint.
Across most datasets, global parameters (those associated with substitution and site models) showed minimal bias, regardless of how the fixed topology was obtained (Figure~\ref{fig:insensitive}).
The HIV dataset was a notable exception, showing consistently large deviations across all fixed-topology analyses.
This dataset was the only one where two parameter estimates fell outside the $\pm 25\%$ relative error range.
These results suggest that for many routine molecular clock analyses, fixed-topology approaches may suffice for substitution and site model estimation.

\begin{figure}[H]
\centering
\includegraphics[width=0.95\textwidth]{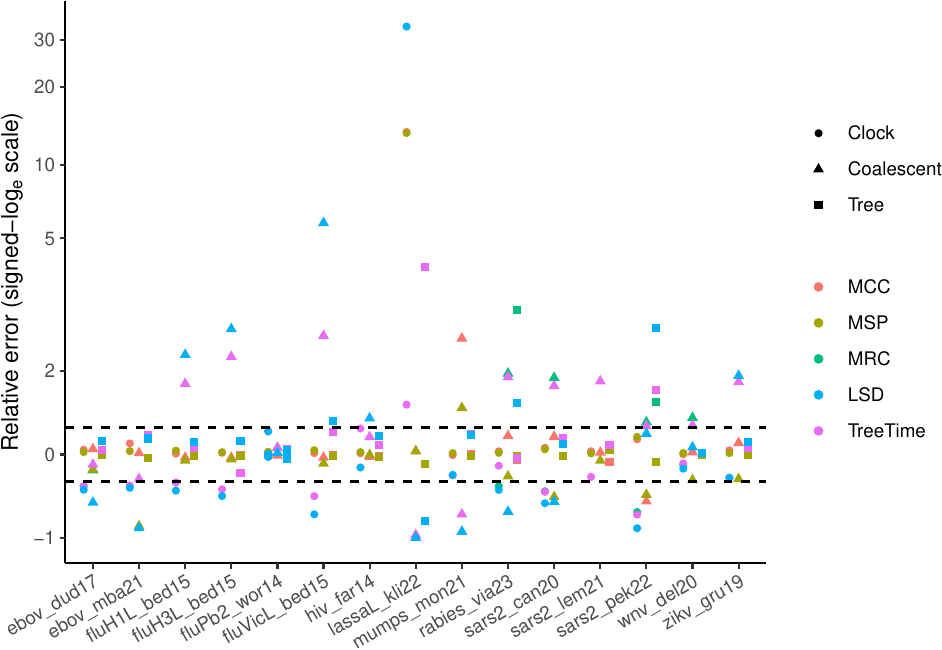}
\caption{\
Relative error for clock and coalescent model parameters, and for the root height parameter of the tree model.
Y-axis plotted on a signed log scale: $\textrm{sign}(y) \log_e(1+|y|)$, with ticks labeled in the original scale.
Results are shown for five fixed-topology methods: maximum clade credibility (MCC), maximum sampled posterior (MSP), maximum rooting credibility (MRC), LSD, and \treetime.
Dotted lines show the -25\% and 25\% relative error thresholds.
}%
\label{fig:sensitive}
\end{figure}

\textbf{Time-dependent parameter estimates are sensitive to the fixed-topology assumption.}
In contrast, parameters that depend on the temporal structure of the tree, including clock rates, node heights (e.g., the root age), and coalescent population size parameters, exhibited greater sensitivity to the fixed-topology constraint (Figure~\ref{fig:sensitive}).
Overall, $75\%$ (189/252) of parameters fell within the $\pm 25\%$ relative error band, with the remainder split evenly between values above and below it.
Notably, $93\%$ of parameters inferred from posterior-informed topologies (MCC and MSP trees) were within the interval, compared with only $62\%$ for trees rooted with \treetime\ or LSD.
Rooting the ML tree with the MRC method substantially improved accuracy, yielding $92\%$ of parameters within the $\pm 25\%$ interval.
Thus, among the fixed-topology approaches, those based on trees sampled from the unconstrained posterior (i.e., the maximum clade credibility tree or the tree with the highest posterior probability) generally resulted in smaller deviations than trees derived via \iqtree/LSD or \iqtree/\treetime, consistent with their higher compatibility with the full Bayesian model.

\begin{figure}[H]
\centering
\includegraphics[width=0.95\textwidth]{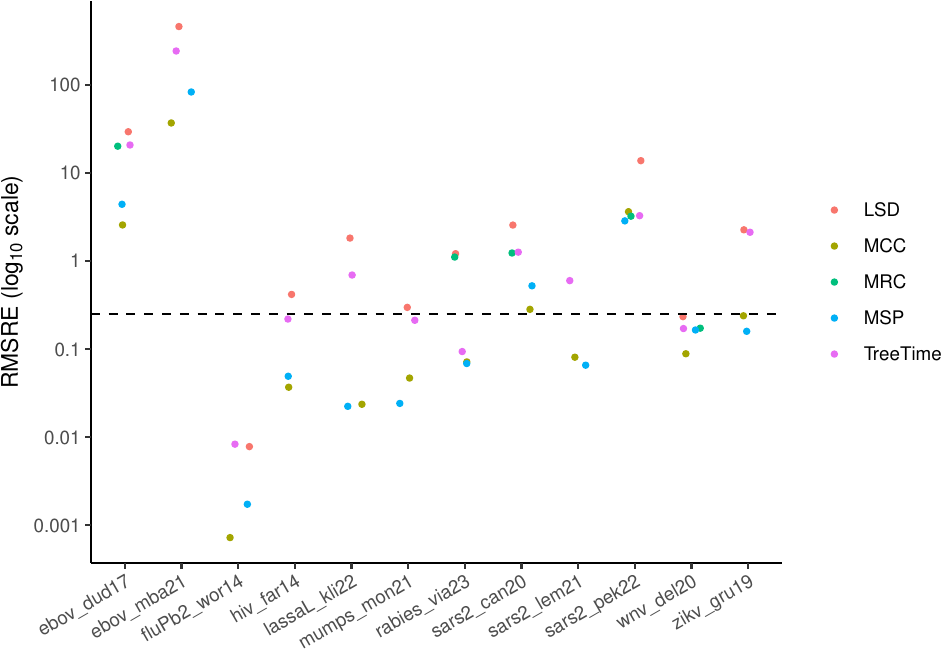}
\caption{\
Root mean squared relative error (log scale) for parameters of the piecewise-constant population size model.
Results are shown for five fixed-topology methods: maximum clade credibility (MCC), maximum sampled posterior (MSP), maximum rooting credibility (MRC), LSD, and \treetime.
Dotted line shows 25\% error.
}%
\label{fig:skygrids}
\end{figure}

This trend also held for the RMSRE of population size estimates (Figure~\ref{fig:skygrids}).
Across datasets, the mean RMSRE of the piecewise-constant parameters was substantially higher for trees rooted with \treetime\ or LSD, averaging nearly six times greater than that obtained from posterior-informed topologies (MCC and MSP).
While 46\% of all parameters were within the $\pm 25\%$ relative error interval, this proportion dropped from 67\% for posterior-informed topologies (MCC and MSP) to only 30\% for trees rooted with \treetime\ or LSD.
In all cases, the MRC method outperformed LSD, and in three out of five datasets it also yielded lower errors than \treetime.

\begin{figure}[h]
\centering
\includegraphics[width=0.95\textwidth]{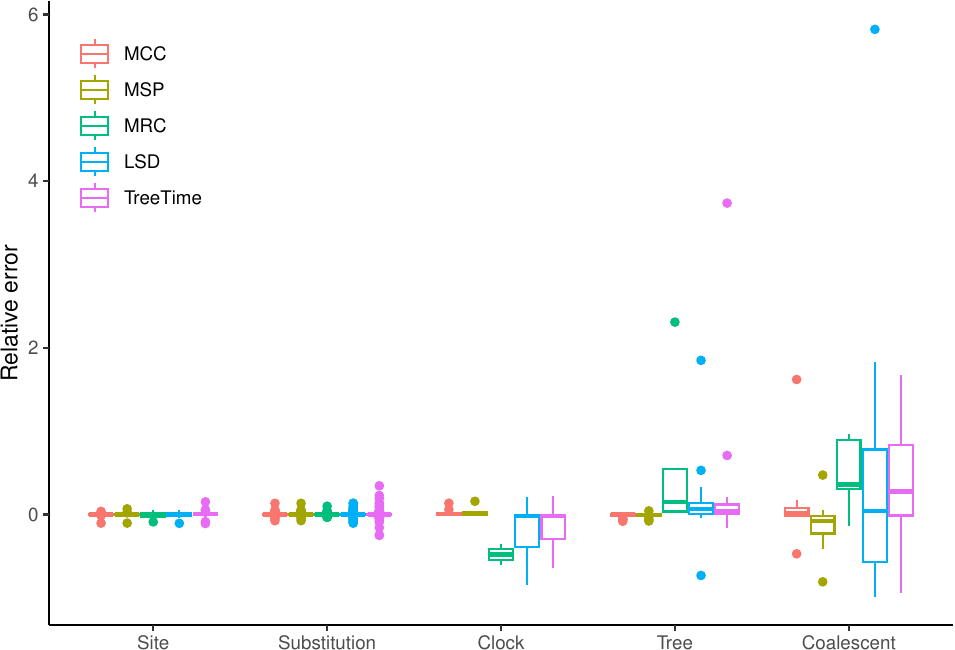}
\caption{\
Relative biases of parameter estimates for fixed-topology analysis relative to an unconstrained BEAST analysis.
Each point represents the relative bias of a parameter estimate for a specific dataset.
Boxplots show the relative bias for each inference method: Maximum Clade Credibility (MCC), Maximum Sampled Posterior (MSP), Maximum Rooting Credibility (MRC), TreeTime, and LSD.
}%
\label{fig:re}
\end{figure}

When all parameter types are viewed together, the contrast becomes clear: substitution and site model parameters are generally well estimated regardless of the fixed topology, whereas clock, tree, and coalescent model parameters show greater sensitivity (Figure~\ref{fig:re}).

\begin{figure}[H]
\centering
\includegraphics[width=0.95\textwidth]{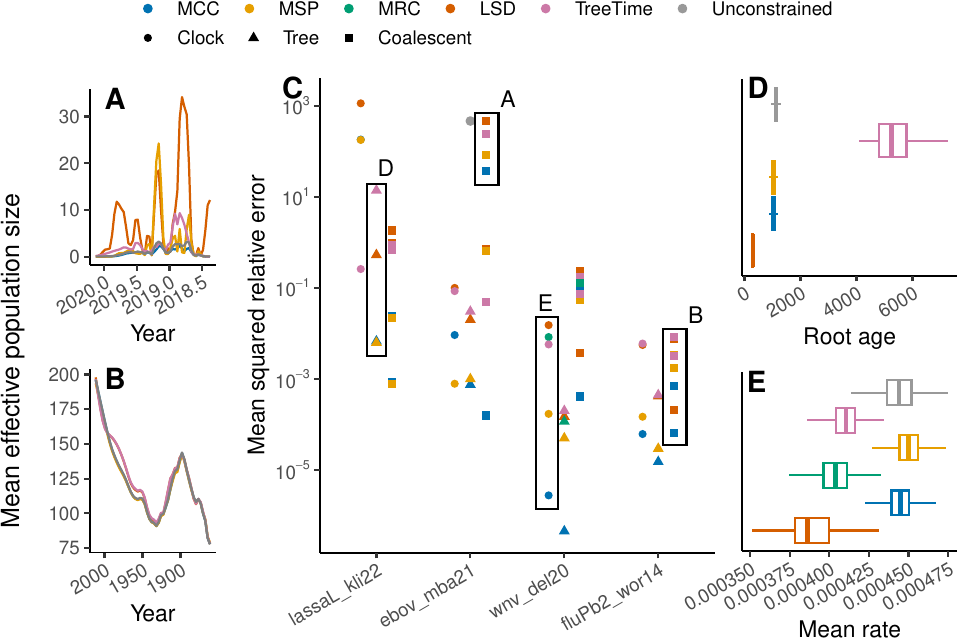}
\caption{\
Mean squared relative errors in clock, coalescent, and tree (i.e. root height) parameter estimates from fixed-topology analyses across selected datasets, using five inference methods: maximum clade credibility (MCC), maximum sampled posterior (MSP), maximum rooting credibility (MRC), \treetime, and LSD.
Panel C shows the main scatter plot of relative errors across methods.
Selected points are highlighted with boxes labeled A, B, D, and E, corresponding to subpanels that provide further detail.
Panels A and B display Skygrid plots for two highlighted datasets.
Panels D and E show the distributions of inferred root ages and mean evolutionary rates, respectively.
}%
\label{fig:subpanels}
\end{figure}

In order to give an idea of what success and failure look like in these tasks, we present the two best and two worst datasets (in terms of root mean squared relative errors) in more detail (Figure~\ref{fig:subpanels}).
Figure~\ref{fig:subpanels}C gives an overview of the comparisons detailed in the other panels.
For the \textsf{ebov\_mba21} dataset, inferring the population size history was challenging when trees were rooted with LSD and \treetime\ (Figure~\ref{fig:subpanels}A).

By contrast, the effective population size through time plots (Figure~\ref{fig:subpanels}B) for the influenza B virus (\textsf{fluPb2\_wor14}) dataset were consistently accurate, independent of the method used to fix the rooted topology.
Figure~\ref{fig:subpanels}D shows that the root age estimates for the Lassa virus dataset (\textsf{lassaL\_kli22}) were particularly poor for \treetime, while Panel E shows that the clock rate estimates for the West Nile virus (\textsf{wnv\_del20}) dataset were substantially biased when using LSD and \treetime\ but not when using posterior-informed topologies.

\noindent\textbf{ML-rooted trees do not always appear in the BEAST posterior.}
To explore the overlap between ML rootings and BEAST, we compared ML-rooted trees generated via \iqtree/LSD and \iqtree/\treetime\ to trees sampled from the unconstrained BEAST posterior.
In several cases, the root selected by LSD and \treetime\ had low posterior support or was never sampled in the posterior (Table~\ref{tab:rooting}).
Importantly, for 10 of the datasets, the splits induced by the root placements in the BEAST tree samples were not present in the ML trees inferred by \iqtree\, indicating that ML unrooted trees were poor surrogates for conducting time-tree analyses under a fixed topology.

\begin{table}[H]
\centering
\begin{tabular}{ |c||c||c|c||c| }
 \hline
Dataset & \# unique root & LSD & TreeTime & \iqtree \\
 \hline
\textsf{ebov\_dud17} & 109 & 0.0 & 0.1116 & \textbf{0.2018} \\
\textsf{ebov\_mba21} & 25 & 0.0 & 0.0 & 0.0 \\
\textsf{fluH1L\_bed15} & 40 & 0.0 & 0.788 & 0.788 \\
\textsf{fluH3L\_bed15} & 3 & 0.0005 & 0.0 & 0.0005 \\
\textsf{fluPb2\_wor14} & 1 & 1.0 & 1.0 & 1.0 \\
\textsf{fluVicL\_bed15} & 7 & 0.0 & 0.9967 & 0.9967 \\
\textsf{hiv\_far14} & 127 & 0.0 & 0.0 & 0.0 \\
\textsf{lassaL\_kli22} & 5 & 0.6794 & 0.6794 & 0.6794 \\
\textsf{mumps\_mon21} & 4 & 0.9977 & 0.9977 & 0.9977 \\
\textsf{rabies\_via23} & 19 & 0.1012 & 0.0358 & \textbf{0.2036} \\
\textsf{sars2\_can20} & 882 & 0.1486 & 0.0384 & \textbf{0.4449} \\
\textsf{sars2\_lem21} & 1107 & 0.0 & 0.0 & 0.0 \\
\textsf{sars2\_pek22} & 4359 & 0.0 & 0.0474 & \textbf{0.3035} \\
\textsf{wnv\_del20} & 3785 & 0.0 & 0.1002 & \textbf{0.1085} \\
\textsf{zikv\_gru19} & 73 & 0.0 & 0.0 & 0.0 \\
 \hline
\end{tabular}
\caption{Summary of rooting information across datasets.
Number of unique root placements in the unconstrained BEAST analyses, based on split support from 10,000 trees.
The LSD and TreeTime columns indicate the proportion of trees from the unconstrained BEAST analyses that share the same root placement as LSD or \treetime, respectively.
The \iqtree\ column contains the highest root placement probability from the unconstrained BEAST analyses compatible with the maximum likelihood \iqtree\ tree.
Bold values indicate an increase in root placement probability compared to LSD or \treetime.}
\label{tab:rooting}
\end{table}

\noindent\textbf{Incorrect root placement introduces systematic biases in parameter estimates.}
To disentangle the effect of rooting from that of topology, we tested whether assigning a more accurate root to the ML tree improved the performance of fixed-tree analyses.
Specifically, we rooted the ML trees using the root placements with the highest posterior probability inferred from the unconstrained BEAST analyses.
This approach yielded only five datasets where a better root was identified compared to \treetime\ and LSD (Table~\ref{tab:rooting}).
Strikingly, for four datasets, the ML tree was incompatible with all rootings sampled in the full BEAST analyses.
Even when improved rooting was possible, only 50 and 52 out of 112 parameters showed better estimates than the LSD- and \treetime-rooted analyses, respectively.
These results suggest that root placement alone is insufficient to achieve the accuracy of unconstrained analyses; instead, the overall topology exerts a stronger influence on parameter estimates.

\begin{figure}[h]
\centering
\includegraphics[width=\textwidth]{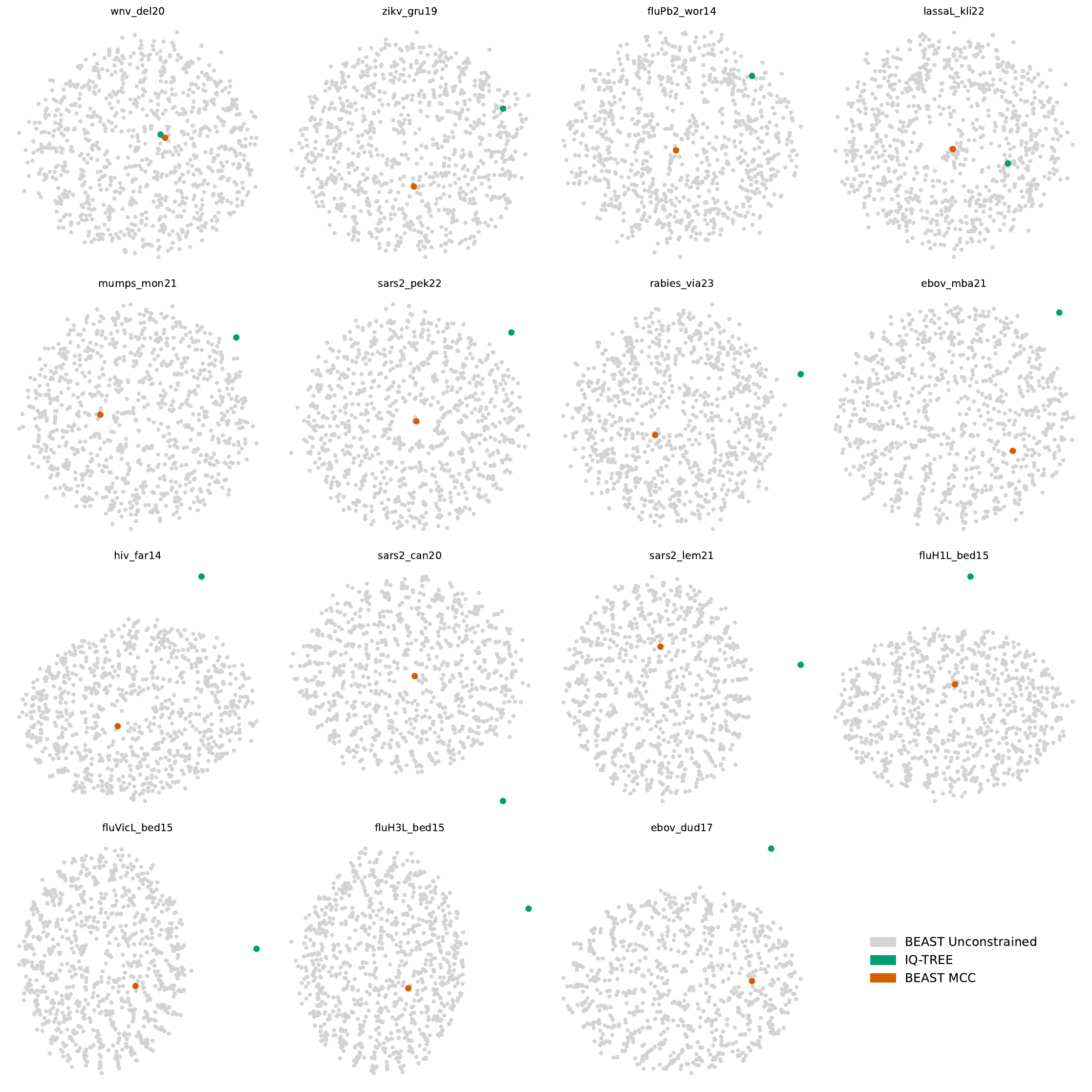}
\caption{\
Two-dimensional graphs showing pairwise distances between unrooted phylogenetic trees inferred from Robinson-Foulds distances.
Each node represents a tree, and the spatial relationships between nodes reflect their pairwise topological distances.
Each graph includes trees sampled from an unconstrained BEAST analysis, the maximum clade credibility (MCC) tree, and the maximum likelihood tree inferred by IQ-TREE.
Graphs are ordered by increasing standardized (z-score) ML--posterior distance (Table~\ref{tab:zscore}).
}%
\label{fig:cluster}
\end{figure}

\noindent\textbf{ML topologies are often incompatible with topologies supported by the unconstrained BEAST posterior.}
A two-dimensional embedding of tree space showed a clear separation between topologies inferred with time-aware models estimated in BEAST and those obtained under unrooted models with \iqtree\ (Figure~\ref{fig:cluster}).
In most cases, the ML tree lay outside the BEAST posterior cluster, but for four datasets (\textsf{wnv\_del20}, \textsf{zikv\_gru19}, \textsf{fluPb2\_wor14}, and \textsf{lassaL\_kli22}) it fell within the cloud, suggesting greater topological similarity.
Notably, whereas the \textsf{wnv\_del20} dataset yielded accurate parameter estimates, the \textsf{lassaL\_kli22} dataset displayed pronounced biases, highlighting that proximity in tree space does not necessarily translate to accurate inference.

\section{Discussion}

In this paper, we have shown that fixing a topology can substantially impact phylodynamic inference, particularly for parameters that are closely tied to the tree structure.

The fixed tree assumption was problematic even when the tree was drawn from the BEAST posterior.
Specifically, substitution and site model parameters were largely robust to the choice of topology, whereas temporal and demographic parameters exhibited notable biases when the topology was fixed.
These findings underscore that the validity of the fixed-topology assumption is highly parameter-dependent.

The most pronounced effects were observed in clock rate and demographic inference.
These parameters rely on the relative ordering and spacing of coalescent and sampling events in the tree, which are highly sensitive to topological uncertainty.
Our results indicate that ignoring this uncertainty---by conditioning on a single tree---can lead to biased estimates that may misrepresent the underlying evolutionary dynamics.
Although posterior-informed topologies (e.g., MCC or MSP trees) mitigated some of these issues, they did not eliminate them entirely.
This reflects the fact that even posterior summary trees cannot fully capture the uncertainty present in the tree space explored during Bayesian inference.

The effects were even larger when using two-step methods such as maximum likelihood tree inference followed by molecular dating using LSD or \treetime.
These approaches assume that an ML tree inferred without temporal information is sufficiently close to the true time-scaled tree to support accurate dating in the second step.
However, our results show that this assumption does not always hold.
When the initial topology is inconsistent with the temporal signal, subsequent dating analyses cannot correct the discrepancy, and the resulting estimates propagate the error.
This highlights a fundamental limitation of the two-step paradigm: temporal and topological information are not separable, and their joint inference is essential for accurate phylodynamic reconstruction.

These results challenge the routine use of fixed topologies in phylodynamic analyses.
Although fixing a topology offers clear computational advantages (particularly when analyzing large datasets or when complex models make joint inference computationally prohibitive), it comes at the cost of potentially misleading parameter estimates.
For some applications, such as substitution model characterization or site-specific rate heterogeneity, this trade-off may be acceptable.
For others, especially where precise estimates of epidemic histories or clock rates are required, the risks associated with fixed topologies are difficult to justify.

A limitation of our study is that we consider the results of unconstrained Bayesian analyses as our gold standard against which we assess the validity of fixed-topology approaches.
Even these unconstrained analyses are themselves subject to model misspecification \citep{gao2023model} and limited mixing \citep{Gao2025-bf}.
Nevertheless, our paper establishes that the frequently used two-step approach to phylodynamics does not result in the same conclusions as a full Bayesian analysis of the joint distribution.

Looking forward, methodological advances are needed to reconcile computational efficiency with statistical rigor.
One promising direction is the development of fast, likelihood-based methods that directly infer time trees, estimating molecular clock rates, node ages, and root positions in a single step.
Incorporating time-aware models into widely used ML frameworks such as \iqtree\ would represent a major advance, potentially rendering current two-step dating pipelines obsolete.
Beyond maximum likelihood, such developments could also provide stronger foundations for approximate Bayesian methods such as variational inference, the accuracy of which depends critically on the quality of the trees it conditions upon.

\section{Acknowledgments}
This project was partially supported by US National Institutes of Health grants R01 AI162611 and R01 AI153044.
Computational facilities were provided by the UTS eResearch High Performance Computer Cluster.
Scientific Computing Infrastructure at Fred Hutch was funded by ORIP grant S10OD028685.
Dr.\ Matsen is an Investigator of the Howard Hughes Medical Institute.

\clearpage

\bibliographystyle{plainnat}
\bibliography{main}

\newpage

\beginsupplement

\section*{Supplementary Materials}

\begin{table}[H]
\centering
\begin{tabular}{ |c|c|c|c|c|c|c| }
 \hline
Dataset & Virus & Taxa & Length & Coalescent & Substitution & Clock \\
 \hline
ebov\_dud17 & EBOV &   1610       &  18992 &  Skygrid & \makecell{ 1:14517[1+2+3(HKY+$\Gamma_4$)] \\ 14518:18992(HKY+$\Gamma$4)} & UCLN \\
ebov\_mba21 & EBOV & 297       &  16757 & Skygrid & 12+3(HKY+$\Gamma_4$) & UCLN \\
fluH1L\_bed15 & IAV & 2144    & 1695 &   Constant & 12+3(HKY) & Strict \\
fluH3L\_bed15  & IAV & 4006    & 1698 &   Constant & 12+3(HKY) & Strict \\
fluPb2\_wor14  & IAV & 354    & 2280 &    Skygrid & 12+3(HKY+$\Gamma_4$) & Local \\
fluVicL\_bed15 & IBV & 1999     & 1755 &   Constant & 12+3(HKY) & Strict \\
hiv\_far14    & HIV  & 927      & 438 &  Skygrid & GTR+$\Gamma_4$ & UCLN \\
lassaL\_kli22 & LASF & 551     & 7038 &  Skygrid & GTR+$\Gamma_4$ & UCLN \\
mumps\_mon21 & MuV & 467 &  15393 &     Skygrid & HKY+$\Gamma_4$ & Strict \\
rabies\_via23  & RABV  & 290     & 11883 &  Skygrid & 12+3(GTR+$\Gamma_4$) & Strict \\
sars2\_can20 & SARS‑CoV‑2 & 1046    & 29409 & Skygrid & HKY+$\Gamma_4$ & Strict \\
sars2\_lem21 & SARS‑CoV‑2  & 3241    & 29409 & Skygrid & HKY+$\Gamma_4$ & Strict \\
sars2\_pek22 & SARS‑CoV‑2  &  717    &    29232 & Skygrid &  GTR+I & Strict \\
wnv\_del20 & WNV    & 801   &  10302  & Skygrid & GTR+$\Gamma_4$ & UCLN \\
zikv\_gru19   & ZIKV  & 283  &     10269  & Skygrid & 1+2+3(HKY+$\Gamma_4$) & UCLN \\
 \hline
\end{tabular}
\caption{Dataset and model specifications.
The HKY+$\Gamma_4$ model applies a common HKY substitution model to all sites, with rate heterogeneity modeled using a discretized gamma distribution with four categories.
The +I model indicates that a proportion of sites are invariant.
In the 12+3(HKY) model, the first and second codon positions share the same HKY substitution model while the third position has a separate HKY model.
In the 1+2+3(HKY) model, each codon position (1st, 2nd, and 3rd) is modeled with its own independent HKY substitution model.
UCLN refers to an uncorrelated log-normal relaxed molecular clock.}
\label{tab:data}
\end{table}

\begin{table}[h] 
\centering
\begin{tabular}{ |c|c| }
 \hline
Dataset & z-score \\
 \hline
wnv\_del20 & -0.91\\
zikv\_gru19 & 0.17\\
fluPb2\_wor14 & 0.88\\
lassaL\_kli22 & 1.20\\
mumps\_mon21 & 1.67\\
sars2\_pek22 & 1.88\\
rabies\_via23 & 2.57\\
ebov\_mba21 & 2.78\\
hiv\_far14 & 3.57\\
sars2\_can20 & 4.19\\
sars2\_lem21 & 4.97\\
fluH1L\_bed15 & 5.34\\
fluVicL\_bed15 & 6.24\\
fluH3L\_bed15 & 6.47\\
ebov\_dud17 & 6.72\\
\hline
\end{tabular}
\caption{Mean ML--posterior distances standardized as z-scores.}
\label{tab:zscore}
\end{table}

\end{document}